\newif\ifsubmode
\submodefalse

\newif\ifprintfig
\printfigtrue

\newcommand\clock{\count0=\time \divide\count0 by 60
     \count1=\count0 \multiply\count1 by -60 \advance\count1 by \time
     \number\count0:\ifnum\count1<10{0\number\count1}\else\number\count1\fi}

\ifsubmode
  \documentstyle[12pt,aasms4,psfig]{article}
  \received{}
  \accepted{}
  \journalid{}{}
  \articleid{}{}
\else
  \documentstyle[11pt,aaspp4,psfig]{article}
  \tightenlines
\fi

\lefthead{Dalcanton \& Hogan}
\righthead{Phase Space Constraints}

\font\cap=cmcsc10

\newcommand\hi{\noindent \hangindent=2.5em}
\newcommand\pc{{\rm\,pc}}

\newcommand\kpc{{\rm\,kpc}}
\newcommand\Mpc{{\rm\,Mpc}}

\newcommand\kmsec{{\rm\,km/s}}
\newcommand\kms{\kmsec}

\newcommand\msun{{\rm\,M_\odot}}

\newcommand\eV{{\rm\,eV}}

\begin{document}

\title{Halo Cores and Phase Space Densities: Observational
Constraints on Dark Matter Physics and Structure Formation}

\author{Julianne J. Dalcanton\altaffilmark{1} \& 
        Craig J.\ Hogan\altaffilmark{2}}
\affil{Department of Astronomy, University of Washington, Box 351580,
Seattle WA, 98195}

\altaffiltext{1}{e-mail address: jd@astro.washington.edu}
\altaffiltext{2}{e-mail address: hogan@astro.washington.edu}
  

\begin{abstract}
  We explore observed dynamical trends in a wide range of
  dark-matter-dominated systems (about seven orders of magnitude in
  mass) to constrain hypothetical dark matter candidates and scenarios
  of structure formation.  First, we argue that neither generic warm
  dark matter (collisionless or collisional) nor self-interacting dark
  matter can be responsible for the observed cores on all scales.
  Both scenarios predict smaller cores for higher mass systems, in
  conflict with observations; some cores must instead have a dynamical
  origin.  Second, we show that the core phase space densities of
  dwarf spheroidals, rotating dwarf and low surface brightness
  galaxies, and clusters of galaxies decrease with increasing velocity
  dispersion like $Q\propto \sigma^{-3}\propto M^{-1}$, as predicted
  by a simple scaling argument based on merging equilibrium systems,
  over a range of about eight orders of magnitude in $Q$.  We discuss
  the processes which set the overall normalization of the observed
  phase density hierarchy.  As an aside, we note that the observed
  phase-space scaling behavior and density profiles of dark matter
  halos both resemble stellar components in elliptical galaxies,
  likely reflecting a similar collisionless, hierarchical origin.
  Thus, dark matter halos may suffer from the same systematic
  departures from homology as seen in ellipticals, possibly explaining
  the shallower density profiles observed in low mass halos.  Finally,
  we use the maximum observed phase space density in dwarf spheroidal
  galaxies to fix a minimum mass for relativistically-decoupled warm
  dark matter candidates of roughly 700$\eV$ for thermal fermions,
  and 300$\eV$ for degenerate fermions.

\end{abstract}



\section{Introduction}                  \label{sec:intro}

Recent work has drawn attention to the apparent conflict between 
predictions of collisionless Cold Dark Matter (CDM) on small scales
and observations of rotation curves of dark matter dominated
galaxies.  Numerical simulations suggest that in a CDM cosmology, dark
matter halos should have steeply rising central cusps ($\rho\propto
r^{-1.5}$) and high densities.  While the observational conclusions
are somewhat ambiguous on the innermost profile shapes (e.g.\ Swaters
et al.\ 2000, van den Bosch et al.\ 2000, van den Bosch \& Swaters
2000, Borriello \& Salucci 2000, Dalcanton \& Bernstein 2000, Burkert
1997), rotation curves consistently imply low characteristic halo
densities in the central regions.  Because $\rho(<r)\propto
(V(r)/r)^2$, the characteristic slope of the rotation curve is
proportional to the square root of the mean enclosed density.
Observations of dark matter dominated galaxies consistently find
rotation curves which rise with $V/R \sim 10-20\kms$, suggesting much
lower characteristic densities than implied by simulations of halos in
viable CDM cosmologies, where $V/R \sim 30-50\kms$ (Moore et al.\ 
1998, 1999, Navarro et al.\ 1996, 1997).

The growing belief that there truly is a conflict between theory and
observations has led to a renaissance in exploring alternative
models for dark matter.  By violating either the ``collisionless''
or ``cold'' properties of traditional CDM, or by considering
additional exotic properties, many authors have sought to preserve the
successes of CDM on large scales, while modifying the manifestations
of dark matter on small scales (Spergel \& Steinhart 1999, Hogan \&
Dalcanton 2000, Sommer-Larson \& Dolgov 1999, Mohapatra \& Teplitz
2000, Peebles 2000, Goodman 2000, Riotto \& Tkachev 2000, Hu et al.\ 
2000, Shi \& Fuller 1999, Colin et al. 2000.)

In this paper, we   place broad constraints upon these
alternatives to CDM, by revisiting observations of the structure and
phase space density of halos over a wide range of scales.  Many of the
above alternative dark matter scenarios make specific predictions for
the sizes of dark matter cores as a function of mass scale (see
Spergel \& Steinhart 1999, Hogan \& Dalcanton 2000, Hannestad 1999,
Burkert 2000, Kochanek \& White 2000, Yoshida et al.\ 2000, Moore et
al.\ 2000, although some of these are in conflict with each other,
particularly regarding the long term stability of self-interacting
cores).  We confront these predictions with existing limits on the
scale of inner halo cores in \S\ref{sec:cores}, and argue that neither
packing of phase space nor highly collisional dark matter can be
primarily responsible for the observed behavior of dark matter cores
at all scales.

In addition to discussing the scaling behavior of dark matter cores,
we focus upon trends of phase space density $Q$ as a probe of
structure formation history.  In \S\ref{sec:evolution} we summarize
the statistical and dynamical behavior of $Q$ in hierarchical
clustering, including a simple argument predicting the decrease of $Q$
with mass and velocity dispersion.  Current observational limits on
the variation of phase space density with mass are reviewed and
summarized in \S\ref{sec:observations}.  We show in \S\ref{sec:interp}
that the phase space density of the dark matter halos is a very
strongly declining function of mass (consistent with the
nearly constant density seen across a similar mass scale in Firmani et
al.\ 2000).  This behavior is exactly as predicted by models
where halos formed hierarchically, with successive mergers
leading to phase mixing and dilution of the coarse-grained
distribution function.  We also note that the observed decline in
phase space density for dark matter halos and their predicted density
profiles resembles that observed in the baryon-dominated central
regions of giant elliptical galaxies, and that by analogy, the
structure of dark matter halos may also suffer from the systematic
departures from homology similar seen in ellipticals.  Finally, we use
the maximum observed phase space densities to   derive a limit on
particle mass of $> 700\eV$ for relativistically decoupled thermal
relics.

\section{Cores in Collisionless and Collisional Solutions}                
\label{sec:cores}

Nonsingular dark matter central halo profiles appear in a wide range
of environments, a fact which already argues against the simplest
explanation of cores based on ``phase packing.''  We illustrate this
point using two such scenarios for limiting the maximum density of
cores in dark matter halos (Hogan \& Dalcanton 2000), namely
generalized warm collisionless dark matter and warm collisional dark
matter.

In the first scenario, dark matter particles have some primordial velocity
dispersion, leading to a ``phase space
density'' $Q\equiv \rho/\langle v^2\rangle^{3/2}$ whose coarse-grained
value is then either preserved or decreased during subsequent epochs
of structure formation (see \S\ref{sec:evolution}, equation
\ref{eqn:entropy}).  For halo material with a roughly isothermal,
isotropic velocity dispersion $\sigma$, the primordial phase density
therefore sets a minimum core size corresponding to ``phase packing''
the material:
\begin{equation}
\label{eqn:r_Q}
r^2_{c,min} = \frac{\sqrt{3}}{4\pi G Q_0}\frac{1}{\left<\sigma^2\right>^{1/2}}.
\end{equation}
If the halo has gone through a period of violent relaxation or shock
heating which thoroughly heats the matter ($Q_0 \rightarrow
Q^\prime<Q_0$ everywhere), then the core may be larger than
$r_{c,min}$.

In the second scenario, the dark matter particles are highly
collisional, and thus they behave as a gas\footnote{In Hogan \&
  Dalcanton (2000), we argued that the moderately collisional case
  would not be stable.  A constant-density core requires a temperature
  gradient for support.  However, if dark matter were only moderately
  collisional, particles would diffuse outwards in less than a Hubble
  time, erasing the necessary temperature gradient.  This diffusive
  heat conduction would runaway to form a dense central cusp.  Thus,
  we restrict ourselves to the highly collisional case, where the
  diffusion time is sufficiently long for cores to be stable over a
  Hubble time.}.  The equilibrium configuration of the halo will
therefore be the solutions of a classical, self-gravitating, ideal
gas.  For a polytropic equation of state ($p\propto\rho^{\gamma}$) at
all radii (i.e. constant entropy), and assuming the system is
non-relativistic and adiabatic ($\gamma=5/3$), the density profile of
the halo becomes that of a Lane-Emden polytrope like a giant
degenerate dwarf star.  (The material here is not degenerate but is on
an adiabat again limited by the initial $Q_0$.)  For a total mass M
and radius R, solutions of the Lane-Emden equation give a central
density $\rho=1.43M/R^3$ and central pressure $p=0.77GM^2/R^4$.  Using
the equation of state, $M\propto R^{-3}$, and thus the characteristic
velocity of the sphere is $\sigma^2=GM/\sqrt{3}R \propto R^{-4}$,
which is a similar scaling to equation~\ref{eqn:r_Q}.

For both of these cases, we predict a simple scaling relationship
between the size of dark matter cores and the characteristic velocity
dispersion of the system.  In general, higher mass, high velocity
dispersion systems should have smaller cores, with $r_{core} \propto 1
/ \sqrt{\sigma}$.  For dwarf spheroidals ($\sigma\!\sim\!10\kms$),
rotating dwarf and LSB galaxies ($\sigma\!\sim\!50-100\kms$), and
clusters ($\sigma\!\sim\!1000\kms$), we expect core sizes to scale
like roughly 10:2-3:1, if in fact the core size is set more by
primordial conditions than by subsequent heating/relaxation.  If we
take a fiducial core size of $1\kpc$ for dwarf spheroidals, then
rotating dwarfs and LSBs would have 300--500$\pc$ cores, and primordial galaxy
cluster cores would be microscopic (100$\pc$) and observationally
undetectable.  Likewise, if we set the fiducial scale at clusters,
with $r_{core}\sim50\kpc$, then dwarf spheroidals would have
implausibly large cores (0.5$\Mpc$, comparable to the separation
between giant spirals in the Local Group, and inconsistent
with possible detections of extra-tidal stars in nearby spheroidals;
e.g. Majewski et al.\ 2000).

Even if we are detecting cores limited by primordial phase space density in
dwarf spheroidals, similar cores in larger systems would be
undetectable.  Detectable cores in more massive galaxies and clusters
must therefore be due to other processes, for example heating and/or
violent relaxation during formation.  This implies that the properties
of warm or self-interacting dark matter would be most directly probed
by the properties of dwarf spheroidals, not rotating dwarf and low
surface brightness galaxies.  On the other hand, if the cores seen in
rotating LSBs are due to the effects of self-interaction, then the
dense halos of dwarf spheroidals may be the end result of core
collapse and may indeed be singular.  We will return to many of these
points below, when we consider the phase space density
of dark matter cores.

\section{Evolution and Scaling of Phase Space Density} 

\subsection{Predictions of Phase Space Density in Gravitational Clustering}
\label{sec:evolution}

We may characterize systems by their mass per volume of phase space, or
``phase space density'' $Q$, a quantity which obeys important
symmetries and in some circumstances admits detailed observational
constraints.  For a collisionless, dissipationless gas, the
fine-grained value of $Q$ does not change, and the evolution of the
system consists of various distortions of the ``phase sheet'' occupied
by particles, in such a way that the coarse grained phase space
density can only decrease. This is related in a straightforward way to
the increase of thermodynamic entropy; for a uniform monatomic ideal 
thermal gas of $N$
particles,
\begin{equation}
\label{eqn:entropy}
S=- kN[\ln (Q)+{\rm constant}].
\end{equation}

The value of the fine-grained phase space density is fixed when the
dark matter particles become microscopically collisionless.  This
quantity $Q_0$ is therefore a primordial relic reflecting the
interactions and masses of the dark matter particles.  Unfortunately,
the primordial value of the fine-grained $Q_0$ cannot be directly
measured;\footnote{In some models $Q_0$ could be measured in a direct
  laboratory detection of dark matter particles, or from effects of
  gravitational lensing of projected catastrophes of surface density
  (Hogan 1999).} the astronomically observable quantity is the mean
coarse-grained phase space density, which can be estimated dynamically
from rotation curves, stellar velocity dispersions, gas emission or
gravitational lensing, using the measured rms velocity and density.
We adopt units for $Q$ most closely related to these observable
quantities: $Q\equiv \rho/\langle v^2\rangle^{3/2}$.  This
coarse-grained phase space density is a strict lower limit to the
primordial $Q$, and we may use observations to set physically
interesting constraints on primordial conditions.

In addition to probing initial conditions, observations of $Q$ can be
made as a function of mass, allowing us to follow the evolution of the
coarse-grained phase space density in hierarchical clustering, i.e.\ 
to study the process of the clustering itself.  We may make a
first order prediction for the expected evolution of $Q$ using the
following simple argument.

Suppose that structures form by hierarchical merging of systems, each
of which is in approximate virial equilibrium. Without gravity or
dissipation, a merged system could be carefully, adiabatically
assembled from parts to almost eliminate any increase in entropy or
decrease of $Q$. Two blobs can be slowly merged into one, and if they
don't mix, the entropy of the new merged blob is just the sum of the
two initial entropies, and therefore $Q$ is preserved. The total phase
space volume is just the sum of the two initial volumes, since nothing
in velocity space changes. The addition of gravitational dynamics to
the picture, however, guarantees that $Q$ decreases steadily as a power
of increasing mass, as the increase in mass requires an increase
in velocity dispersion to maintain virial equilibrium.

We may place limits on the minimum possible decrease in $Q$ which is
compatible with maintaining virial equilibrium during the merging
hierarchy.  Consider the merger of a blob 1 and a
smaller blob 2 into a third blob 3, with the phase density of each
$Q_i=M_i/{\cal V}_i$ depending on the volume ${\cal V}_i$ occupied in
phase space.  All three blobs have a homologous structure and are in
virial equilibrium with characteristic size $R_i$ and velocity
dispersion $\sigma_i$. Set $M_2=\epsilon M_1$, hence
$M_3=(1+\epsilon)M_1$.  We assume that as the small blob 2 sinks into blob 1,
its material is tidally stripped.  The stripping of material with
density $\rho_2$ occurs at a radius where $\rho_1\sim\rho_2$, and
thus this gentle merging process approximately preserves the physical
space density as each layer is homologously added to form the larger
system.  Empirically, the density of dark matter halos is indeed observed
to be approximately constant on mass scales from rotating
dwarfs to clusters (Firmani et al.\ 2000), and thus this gentle merging
assumption may not be a terrible deviation from the truth. 

With our assumption of constant space density during merging, the
added material leads to an increase in volume $R_3=(1+\epsilon)^{1/3}
R_1$.  However, in order to preserve virial equilibrium
$\sigma^2\propto M/R$ it is also necessary to grow the size of the
blob in velocity space, $\sigma_3=(1+\epsilon)^{1/3}\sigma_1$.  Thus,
the system responds by a symmetrical fractional increase in $R$ and
$\sigma$.  The phase volume increases by the factor ${\cal V}_3 =
R_3^3\sigma_3^3 = (1+\epsilon)^2{\cal V}_1$, implying that systems
assembled from this hierarchy follow ${\cal V}\propto M^2$, and hence they
obey\footnote{Note that this is a fundamentally different assumption
  than made by Hernquist et al.\ 1993 for a similar phase space
  evolution calculation.  Following Hausman \& Ostriker 1978, and
  assuming parabolic orbits for merging of identical galaxies, they
  argued that the velocity dispersion of the merged remnant is
  identical to the velocity dispersions of the progenitors, and thus
  that the gravitational radius of the virialized remnant must
  increase in proportion to the mass. This implies that the both the
  physical density and the phase density of the remnants should
  decrease as $M^{-2}$. We argue that this is not the quietest
  limiting hierarchy, although some such mergers certainly occur.}
\begin{equation}      \label{eqn:qsigma}
  Q\propto M^{-1}\propto \sigma^{-3}\propto R^{-3}. 
\end{equation}
\noindent No matter how gradually the assembly is done a certain amount of extra
phase wrapping is needed to achieve this equilibrium, decreasing $Q$.

While not rigorously proved, we argue that equation \ref{eqn:qsigma}
must be close to the slowest possible decline in the coarse-grained
phase space density with increasing mass.  We have assumed the
quietest form of merging, invoking only enough phase mixing to bring
the system into virial equilibrium.  Other models for the evolution of
$Q$ may be invoked (e.g.\ Hernquist et al.\ 1993), but in general
these should involve more phase-mixing, and thus more sharply
declining values of $Q$.  For example, although the predicted scaling
for $Q$ is the same as equation \ref{eqn:qsigma} even when the two
masses are comparable (i.e.\ $\epsilon$ is not much less than one), in
such a situation one expects ``violent relaxation'' rather than tidal
stripping, leading to additional phase mixing and steeper evolution in
$Q$.  Likewise, while the assumption of homology has not been
justified in detail, and indeed no account has been taken of changes
in density profile shapes and other degrees of freedom available to
real systems, these extra degrees of freedom must come with a net
lowering of coarse-grained $Q$ relative to the quiet, homologous case
used in our derivation.  A shallower decrease than equation
\ref{eqn:qsigma} would require the hierarchical merging to create a
systematic increase in physical density --- a situation which we
consider unlikely.  We conjecture that the simple constant-density
scaling may have a rigorous dynamical basis as a limiting
case.

As an aside, we note that similar constant space density behavior may
exist for systems which formed via monolithic top-hat collapse,
instead of through the merging hierarchy.  For very low mass halos
formed from CDM-like power spectra, the characteristic overdensities
$\delta\rho/\rho$ are roughly constant over a range of mass scales,
leading to similar collapse epochs and similar final halo densities.
This will naturally lead to a $Q\sim M^{-1}$ scaling indistinguishible
from the merging hierarchy.  Thus, if these low mass systems survive
and retain the densities imprinted at formation they could be
indistinguishable from the rest of the merging hierarchy.

\subsection{Observational Constraints on Phase Space Densities} 
\label{sec:observations}

We have argued above that considerable information on primordial
conditions, dark matter physics, and the merging hierarchy may be
contained in the scaling of $Q$ with mass.  With this in mind, we now
consider the most recent observational data and derive the best
current measurements of the phase space density of dark matter on mass
scales from $\sim\!10^8\msun$ to $\sim\!10^{15}\msun$.  We will discuss the
interpretation of these results in \S\ref{sec:interp}.

\subsubsection{Phase Space Densities in Dwarf Spheroidals} \label{sec:dwarfsphere}

The lowest mass systems which can be used to measure dark matter
densities are the dwarf spheroidal galaxies in the Local Group.  These
galaxies are extremely diffuse and low mass, and are supported by
velocity dispersion rather than rotation.  They have typical stellar
velocity dispersions of order $10\kms$ and luminosities $\sim\!10^4$
times fainter than bright spiral galaxies (see review by Mateo 1998).
Dynamical mass-to-light ratios for these systems in some cases are
very high ($M_{tot}/L_V\!\sim\!100$), suggesting that they are
completely dark matter dominated\footnote{Alternatively, dwarf
  spheroidals may not be in virial equilibrium, and instead their high
  velocity dispersions may be due to tidal disruption (e.g.\ Kuhn \&
  Miller 1989).  However, there is a relatively tight relationship
  between $M/L$ and luminosity (Mateo et al.\ 1999), suggesting that
  the values of $M/L$ are intrinsic to the galaxy and are not a
  product of environment.  Further support for the existence of dark
  matter in dwarf spheroidals comes from: simulations by Oh et al.\ 
  (1995) which show that dwarfs retain their equilibrium velocity
  dispersion even when being tidally disrupted; Burkert's (1997)
  calculations that the properties of extra-tidal stars in Sextans are
  consistent with a high dark matter content; and Mateo et al's (1998)
  arguments that Leo I (which has one of the highest values of $M/L$)
  is sufficiently isolated that it can not have been strongly affected
  by tidal heating.  While there are clear cases of tidal disruption
  (i.e. Saggitarius), we consider it to be unlikely that tides are
  universally responsible for the high velocity dispersions in dwarf
  spheroidals.}.

We have derived the dark matter density of the dwarf spheroidals
assuming that the dwarfs are dark matter dominated, and that the stars
effectively behave like test particles in the halo potential.  The
dark matter is not required to have the same structure as the stars,
given that current observations are not quite sufficient to distinguish
between the cases of mass following light and of the dwarves being
embedded in a larger halo (e.g.\ Klenya et al.\ 1999).  However,
the possible detection of ``extra-tidal'' stars in Carina (Majewski et
al.\ 2000, Ibata \& Hatzidimitriou 1995, Kuhn et al.\ 1996) beyond a
clear break in the profile at a radius of $\sim30\arcmin$, suggests that
the dark matter core radius is unlikely to be more than a factor of 2
larger than the observed core radius for this particular case.

Assuming that the stars have an isotropic velocity dispersion and
following Pryor \& Kormendy 1990, the central density of the halo is
$\rho_{0D}=\frac{3\ln{2}}{2\pi} \frac{\sigma_*^2}{Gr_c^2}$, where
$\sigma_*$ is the observed one-dimensional central velocity dispersion
of the stars and $r_c$ is the observed ``core'' radius where the
surface density of stars falls to half the central value.  The
corresponding phase space density of the halo of dwarf spheroidals is
therefore
\begin{equation}
Q_{DS} \approx \frac{\rho_{0D}} {(3\eta_*^2\sigma_*^2)^{3/2}} = 
        \frac{\ln{2}}{3^{1/2}2\pi} \frac{1}{G\eta_*^3r_c^2\sigma_*}
\end{equation}
where the scaling factor $\eta_*\equiv\sigma/\sigma_*$ accounts for
the fact that the dark matter particles do not necessarily have the
same velocity dispersion as the stars (which formed from presumably
dissipative baryonic processes).  For an isothermal model for the dark
matter halo, $\rho_{0D}=\frac{9}{4\pi G}\frac{\sigma^2}{r_0^2}$ (where
$r_0$ is the King radius), suggesting that $\eta_* =
0.48\frac{r_0}{r_c}$. We will take $\eta_*\!\sim\!1$, and thus
implicitly assume that the core radius of the dark matter is roughly
twice that of the stellar surface density profile.  Note that because
of the current inability to trace the density profiles of dwarf
spheroidals, $Q_{DS}$ is not necessarily a central phase space
density, but instead is representative of the mean phase space density
within a core radius.

In calculating $Q_{DS}$, we have used data for the eight Local Group
dwarf spheroidals from Mateo (1998) which are fainter than $M_V=-14$
and which have internal kinematic measurements, excluding the tidally
disrupting dwarf Saggitarius.  We have used values for $\sigma_*$ and
$r_c$ given in the compilation of Mateo (1998), but have supplemented
these with newer values for Ursa Minor and Draco from Klenya et al.\ 
(1999) and for Leo I from Mateo et al.\ (2000).


\subsubsection{Phase Densities in Rotationally Supported Galaxies}
\label{sec:rotdwarf}

On somewhat larger mass scales than the dwarf spheroidals,
the stars and gas in galaxies tend to become rotationally supported.
Measurements of the rotation speed as a function of radius $V_c(r)$
can be used to derive the mass interior to $r$, assuming that the
disk is in centrifugal equilibrium.  Assuming that the dark matter
halo is spherical and dominates the mass at all radii, we can
approximate the mean density of the halo within a radius $r$ as
\begin{equation}                        \label{eqn:rhogal}
\rho_{gal}(<\!r) \approx \frac{3}{4\pi G} \frac{V^2_c(r)}{r^2}.
\end{equation}

There is a growing body of evidence that the assumptions which go into
the above equation are not strictly true.  For example, recent
measurements from the flaring of HI disks, the shapes of x-ray
isophotes, warps in galactic disks, and the dynamics of polar ring
galaxies all suggest that galaxy halos are not spherical, but are
somewhat flattened (i.e.\ oblate; see the summary figure in Olling \&
Merrifield 1999, and discussion in Sackett 1999).  However, models by
Olling (1995) show that for the observed range of flattenings, the
true central density is not more than a factor of $\sim\!50$\% greater
what would be derived in the spherical case.

A much larger uncertainty comes from the contribution which the
baryons in the disk make to the dynamics.  The mass of atomic gas is
usually easily determined through the distribution of HI (including a
correction for helium).  However, the molecular gas phase can be the
dominant contributor in the inner disk of massive spiral galaxies, but
is rarely observed, and then is only detected indirectly through the
CO tracer.  The mass in stars is also uncertain, given that for most
stellar populations, most of the mass is due to stars which make
little contribution to the total light.  Thus, the density given by
equation \ref{eqn:rhogal} represents an upper limit to the enclosed
density of the dark matter halo.  Fortunately, for many rotating
dwarfs and low surface brightness galaxies, the baryonic disk is
sufficiently diffuse that for all reasonable stellar mass-to-light
ratios, the enclosed mass is dominated by the dark matter halo.
Unfortunately, this limits our analysis to galaxies with relatively
low rotation speeds ($V_{c,max} \lesssim 100\kms$, vs $V_{c,max} \!\sim\!
250\kms$ for bright spirals).  While some ``Malin-like'' low surface
brightness disks are known to have higher rotation speeds, these
galaxies typically have large central bulges as well, and thus are
likely to be baryon dominated in the central regions.

To calculate $\rho_{gal}$, we have chosen to use the rotation curve
decompositions for NGC 247 ($V_{c,max}\!\sim\!100\kms$; Carignan \&
Puche 1998), DDO 154 ($V_{c,max}\!\sim\!45\kms$; Carignan \& Beaulieu
1989), and NGC 3109 ($V_{c,max}\!\sim\!60\kms$; Jobin \& Carignan
1990), as compiled and analyzed in van den Bosch et al.\ (2000).
These were three cases found in the literature where the HI
observations were sufficiently resolved to accurately trace the
rotation curve in the inner halo.  In this analysis, the core radius
$r_0$ is taken as the radius where the density profile of the dark
matter halo changes slope: $\rho(r) \propto r^{-\alpha}(r + r_0)^{\alpha-3}$

To calculate the phase space density, we derive the halo velocity
dispersion by considering the circular velocity measured at the core
radius.  For an isothermal distribution, the core radius is
$r_0=\sqrt{9\sigma^2/4\pi G \rho}$, suggesting that $V^2_c(r_0)
\approx 3\sigma^2$.  Note, however, that the density distribution of
the halo is not necessarily that of an isothermal sphere, and thus
that our approximation of $\sigma$ will necessarily be uncertain.  For
NGC 3109 and NGC 247, the core radius derived by van den Bosch et al.\ 
(2000) is greater than the last measured point in the rotation
curve\footnote{One complication is that for NGC 247, the core does not
  seem to be constant density (unlike NGC 3109 and DDO 154), and
  instead has a density rising inwards like $r^{-1}$.}, so we take
$\sigma=V_{c,max}/\sqrt{3}$ for these two cases.  For DDO 154, the
best fit core radius is $3\kpc$, well within the last measured radius.
The final velocity dispersions are then $\sigma_{247}=62\kms$,
$\sigma_{3109}=38\kms$, and $\sigma_{154}=22\kms$.  However, because
of the difficulty in securely identifying the halo core radius, these
velocity dispersions are probably uncertain by $\sim\!50$\%.

The resulting average phase space density within $r$ is therefore
\begin{equation}
  Q_D(r) \lesssim \frac{9\sqrt{3}}{4\pi G r^2} \frac{1}{V_c(r_0)} 
                \left(\frac{V_c(r)}{V_c(r_0)}\right)^2.
\end{equation}
%


\subsubsection{Cluster Phase Space Densities}  \label{sec:clust}

Clusters of galaxies provide the best laboratories for measuring the
phase space density of collapsed dark matter halos on the largest mass
scales.  Analyses of their x-ray properties and internal dynamics show that
many (though not all) clusters are well-relaxed systems, and thus
are appropriate for studying the equilibrium state of high-mass
dark matter halos.  

On the other hand, clusters are far from ideal.  Unlike dwarf
spheroidals, the centers of clusters have a significant mass
contribution from baryons, mostly in the form of hot x-ray emitting
gas.  Current estimates are that 10-25\% of the cluster mass within
$r_{500}$ is in the gas phase (c.f.\ Ettori \& Fabian 1999, where
$r_{500}$ is the radius within which the density constrast is 500).
This problem only intensifies in the very centers of clusters, where
the density of the intracluster gas is the highest, and where the
dynamical mass may be dominated by the stellar population of a central
giant elliptical galaxy.  There are also fewer dynamical probes in the
centers of clusters, simply due to limited sampling volume for
velocity tracers.  The best dynamical studies to date, which
incorporate velocity anisotropy for an ensemble of clusters, do not
probe much within a radius of $\sim\!  50h^{-1}\kpc$ (Carlberg et al.\ 
1997, van der Marel et al. 2000).  The mass profiles of the centers of
clusters are also difficult to probe with x-rays.  The historically
low resolution of x-ray telescopes has not allowed measurements of the
temperature of the gas to be spatially resolved at very small scales
(though the experimental situation is rapidly improving).  Finally,
clusters are among the most massive bound structures seen today, and
thus many are still in the process of formation.  We therefore may
expect to see some degree of variation in their properties, reflecting
incomplete relaxation.

Of all the methods of constraining the central densities of clusters,
the most secure estimates come from observations of strong
gravitational lensing within cluster cores (i.e. arcs).  For a
spherical mass distribution, the mean density $\rho_{arc}$ within the
radius of the lensed arc $r_{arc}$ is
\begin{equation}
\rho_{arc} = \frac {3c^2}{16\pi G} (\frac{D_s}{D_l D_{ls}}) \frac{1}{r_{arc}}.
\end{equation}
The use of elliptical mass distributions can reduce this density by
typically 20\%.  To account for the baryonic contribution within
$r_{arc}$, we reduce the above measurement of $\rho_{arc}$
by a factor $(1-f_{baryon})\!\sim\!0.8$ to estimate the
dark matter density.

The resulting average phase space density within $r_{arc}$ is therefore
\begin{equation}
Q_C \approx \frac{(1-f_{baryon})\rho_{arc}} {(3\eta_{gal}^2\sigma_{gal}^2)^{3/2}}.
\end{equation}
We have again included a scaling factor $\eta_{gal}$ to allow for
systematic differences between 1-d velocity dispersion of the galaxies
($\sigma_{gal}$) and the dark matter dark matter.  However, weak
lensing and dynamical studies (c.f.\ Tyson et al.\ 1998, Carlberg et
al.\ 1997) all suggest that mass traces light on $>\!100\kpc$ scales
within clusters, and thus $\eta_{gal}\approx1$.

To calculate $Q_C$ for specific clusters, we have restricted ourselves
to lensing clusters which also host strong cooling flows.  Allen
(1998) argues persuasively that cooling flow clusters are the most
likely to be fully relaxed, and demonstrates that they are the only
clusters whose x-ray and lensing mass estimates are consistent.  We
have plotted $Q_c$ for the six cooling flow clusters with giant arcs
which were analyzed in Allen (1998).  For the three cases where more
detailed mass models have been developed to fit the strong lensing
data, we have revised the spherical estimate of $\rho_{arc}$
accordingly (MS2137.3-2353, Mellier et al.\ 1993; PKS0745-191, Allen
et al.\ 1996; Abell 2390, Pierre et al.\ 1996).

Our calculation for $Q_C$ gives the mean phase space density with the
radius of strongly lensed arcs.  However, we wish to compare $Q_C$
with the phase space densities derived for dwarfs and rotating disks.
These latter quantities are calculated within a core radius, and thus
there is some inherent uncertainty in treating $Q_C$ as a ``core''
phase space density.  At the radii where these arcs are typical,
detailed dynamical studies by Carlberg et al.\ (1997) and van der
Marel et al.\ 2000) show that the cluster density profiles are still
rising like $r^{-1}$ at the innermost measurable radii
($\sim\!35h^{-1}\kpc$); if this behavior continues towards the center
where it terminates in a smaller constant density core, then the core
phase space density could be a factor of 10-100 times higher than the
above estimate of $Q_C$ on scales of $\sim1\kpc$.  Work by Williams et
al.\ (1999) argues that shallow inner cores must be
rare but on the other hand, Tyson et al.\ 1998 reconstruct the
density profile one cooling-flow cluster with a constant density core
($r_c\sim35h^{-1}\kpc$, derived from fitting eight images of a
multiply lensed background galaxy), suggesting that in some cases
$Q_C$ may be close to the true core phase density.  If clusters do not
have a smaller constant density core within $r_{arc}$, and instead the
most appropriate assignment for $r_{core}$ is the typically larger
radius where the density profile changes from $r^{-1}$ to $r^{-3}$,
then the phase space density which should be compared to $Q_{DS}$ and
$Q_{D}$ will be substantially smaller than calculated for $Q_C$.
Considering these uncertainties, our calculated values of $Q_C$ are
uncertain by possibly as much as a factor of 10.

\section{Interpretation}                \label{sec:interp}

We have plotted the characteristic phase space densities for dwarf
spheroidals, rotationally supported galaxies, and clusters of galaxies
in Figures~\ref{fig:Q_vs_V} \& \ref{fig:Q_vs_R}.  It is immediately
apparent that there is a systematic decrease in the phase density as a
function of scale, with more massive systems having dramatically lower
phase space densities\footnote{While the cluster measurements of $Q$
  are uncertain, they are unlikely to compensate for the observed
  factor of $10^8$ variation in phase density.}.  A similar trend was
noted by Burkert (1995) and Sellwood (2000), although over a much
smaller range in scale.

\subsection{Hierarchical Assembly}       \label{sec:hierarchy}

Immediately, the factor of $10^2-10^3$ difference in phase density
between dwarf spheroidals and rotating galaxies in
Figures~\ref{fig:Q_vs_V}~\&~\ref{fig:Q_vs_R} suggests that the cores
in rotating dwarfs cannot be due to ``phase packing'' of material with
a primordial phase density.  This agrees with our analysis in
\S\ref{sec:cores} about the behavior of core sizes as a function of
velocity dispersion.  

Figure~\ref{fig:Q_vs_R} also demonstrates that at the same physical
scale ($\sim 1\kpc$), the phase space densities of rotating dwarfs are
substantially smaller than for dwarf spheroidals\footnote{As a caveat,
  the amount of beam smearing in the inner parts of the rotating dwarf
  measurements may be considerable, leading to artificially low
  central densities (van dan Bosch et al.\ 2000).  These three cases
  have been chosen to minimize this concern, however.} This suggests
that the cores of more massive halos cannot be made up purely of
``sinking satellites'' (e.g.\ Syer \& White 1998); most accreted objects
must have undergone substantial disruption and phase mixing while
being incorporated.

Aside from the above two points, Figures~\ref{fig:Q_vs_V} \&
\ref{fig:Q_vs_R} strongly suggest  that halos formed as the
result of a merging hierarchy.  As discussed above, if larger systems
build up from smaller chunks, relaxation processes necessarily lead to
substantial phase mixing while reaching virial equilibrium, and thus with each
successive merger there is a dilution of the coarse-grained phase
space density.  Figures~\ref{fig:Q_vs_V} \& \ref{fig:Q_vs_R} are
strong support (if any were needed) for hierarchical formation of
galaxies and clusters (i.e. bottom-up, rather than top down); if less
massive objects were to fragment from larger mass objects, their phase
space density would be lower, not higher as is observed, unless there
were substantial dissipation involved in the fragmentation.

Moreover, the behavior seen in Figure~\ref{fig:Q_vs_V} is close
to the minimal decrease $Q\propto 1/\sigma^3$ predicted from our
simple scaling argument in \S\ref{sec:evolution}.  It is a profound
fact that dark matter dominated systems obey a scaling relatively
close to this limit, suggesting that they formed in a fairly quiet
collisionless hierarchy.\footnote{Since we can only measure $Q$ where
  there are stars and gas (i.e.\ in the central regions), it is not
  surprising that they sample the lowest entropy envelope of the
  hierarchy.} The maximum $Q$ on all scales remembers the initial $Q$
at the start of the hierarchy (the top left of
Figures~\ref{fig:Q_vs_V} \& \ref{fig:Q_vs_R}); the $Q$ of the first
systems to collapse sets a maximum $Q$ for all the systems which form
from subsequent clustering.  We discuss the origin of this ``seed point''
in \S\ref{sec:seed} below.

While we believe that cores were built hierarchically on scales from
galaxies to clusters, it is possible that lower mass halos resulted
from early monolithic collapse.  Remarkably, such systems would still
follow the $Q\propto\sigma^{-3}$ scaling.  For small mass halos formed
from a CDM-like power spectra, the density contrast
$(\delta\rho/\rho)_M^2\propto k^3P(k)^2$, is nearly constant with mass,
and thus all perturbations collapse at the same time. The synchronous
formation of low mass halos will lead them to have similar densities,
(if the densities in the final virialized halos tend to track
the density of the universe at their formation -- e.g.\ as claimed by Navarro
et al.\ 1996,1997), yielding the same
$Q\propto\sigma^{-3}$ scaling derived for ``quiet'' merging.  If these
monolithically collapsed systems survive till the present, or if they
subsequently merge, their descendents will be indistinguishable from
the rest of the $Q$-$\sigma$ hierarchy.


We note that the observed scaling of $Q$ is not expected for highly
collisional dark matter.  The prediction of $Q\propto\sigma^{-3}$ was
derived assuming that merging material settles where its density
matches the density of the enveloping galaxy.  In contrast, highly
collisional dark matter is compressed to higher density as it responds
to higher local pressure, then sinks to where it reaches local
pressure and density equilibrium, and finally stops when the entropy
matches.  Thus matter in a quiet merger tends to be stripped at
constant $Q$ rather than constant density. This preserves $Q$ during
merging, leading to $Q\propto\sigma^0$, i.e. constant phase space
density at all mass scales.  An exception occurs if collisions are
rare enough to allow rapid heat conduction, which leads to a
core-collapse instability and high central densities.  This case is of
course dissipational, allowing the $Q$ of a fluid element to increase.
It is possible (though not likely) that the high $Q$ of the low mass
systems arises this way.
 
\subsection{The Seed Point of the Phase Space Hierarchy}   \label{sec:seed}

While the arguments given in \S\ref{sec:evolution} set the slope of
the $Q$-$\sigma$ relation, the origin of the overall normalization
remains unclear.  Within the framework of a merging hierarchy, the
normalization seen in Figure~\ref{fig:Q_vs_V} is fixed by the phase
space density of the lowest mass systems -- the ``seeds'' of the
merging hierarchy -- lying somewhere upwards and to the left on
Figure~\ref{fig:Q_vs_V}.  These early systems will have the highest
observed phase space densities, and will set the phase space density
of all systems further down the merging hierarchy\footnote{Note that
  in CDM there is no upper limit to $Q$ and dark matter halos are
  predicted to exist along a continuation of the $Q(\sigma)\propto
  \sigma^{-3}$ relation indefinitely to low mass.  Moore et al.\ 
  (1999) have speculated that it is this cold initial phase space
  density of CDM which leads to the very dense cores in numerical
  simulations, regardless of the power spectrum of initial
  perturbations.  It is possible, however, that extra heating could be
  provided by smaller-scale dynamical effects not yet resolved.}.
Thus the phase space density of the first generation of collapsed
objects fixes the normalization of the entire $Q$-$\sigma$ relation.
What sets the masses of these early halos and more importantly, what
sets their phase space densities?

There are three physical parameters which can alter the phase space
density of the first collapsed objects.  First is the dark matter
particle microphysics, which sets a maximum value for the fine-grained
primordial $Q_0$, via equation~\ref{eqn:Q_mx} below; no collapsed dark
matter halos can have a coarse-grained phase space density higher than
the primordial value of $Q_0$, unless the dark matter is
dissipational.  Second is the density of the universe during the epoch
when matter first collapses and virializes, if lower initial densities
lead to lower phase densities in the virialized halos (Navarro et al.\ 
1996,1997).  Third is the efficiency\footnote{By ``efficiency'', we
  mean both the fraction of matter which undergoes phase mixing, and
  the degree to which that matter is mixed.} of phase mixing during
collapse itself.  Through violent relaxation and phase-wrapping, the
primordial phase sheet is mixed to a lower coarse-grained phase space
density.  The process of relaxation and virialization during the first
halos' collapse can set the phase density of the lowest mass cores,
and thus the normalization of the entire $Q$-$\sigma$ relation.

The phase space density of the first collapsed objects will be set by
a combination of these three processes.  For example, if violent
relaxation and phase mixing is inefficient (meaning, that much of the
matter is not phase wrapped and is left at high $Q$), and/or the
velocity dispersion of the first collapsed halos is comparable to the
primordial velocity dispersion of the dark matter, then the maximum
observed phase space density will be fixed at the primordial
fine-grained value $Q_0$.  Alternatively, if the original phase sheet
with density $Q_0$ is well phase mixed during collapse of the first
objects, then the maximum observed phase space density will be diluted
from $Q_0$ to a lower value.  Or finally, if the overall density of
the universe is lower when the first halos form, then the early halos
will possibly virialize to lower space densities and lower $Q$.  In
either of the these three cases, at fixed mass (or velocity
dispersion) the seeds for the merging hierarchy will have lower phase
densities than CDM, and the normalization of the $Q$-$\sigma$ relation
will be reduced.  On the other hand, CDM simulations by Moore et al.\ 
(1999) show that introducing an arbitrary cutoff in small-scale power
produces no detectable changes in the density profiles of the dark
matter halos.  This raises questions about the degree of coupling
between the final and initial densities of collapsing halos; it may
turn out that the epoch of collapse has little direct impact on $Q$
in the cores of the first halos.

These three contributors to the $Q$-$\sigma$ normalization --
primordial $Q_0$, violent relaxation \& phase mixing, and the density
at the epoch of collapse -- are not necessarily independent.  In some
cases, the latter two are coupled directly or indirectly to $Q_0$.
For example, the physics of the dark matter sets the primordial phase
space density $Q_0$ via equation~\ref{eqn:Q_mx} below.  This initial
phase space density corresponds to a characteristic velocity
dispersion for dark matter in the early universe.  Given that the
velocity dispersion of the initial conditions may affect the degree of
violent relaxation and the growth of angular momentum, it is possible
that the efficiency of phase mixing during violent relaxation may be
indirectly set by the primordial $Q_0$; in other words, dark matter
particle physics may be imprinted upon the normalization of the
$Q$-$\sigma$ relation, even if $Q_0$ does not fix the phase space
density of the first halos directly.  Likewise, the epoch at which the
first objects collapse depends upon the power spectrum of initial
fluctuations, a function which is in turn affected by $Q_0$.  In warm
dark matter models, for example, free streaming creates a mass
filtering scale $M_{filter}$ (which can be related to $Q_0$), that
suppresses the power spectrum at low masses, and delays the collapse
of the first objects.


If the first objects which collapse have sufficiently low mass, we
expect little change in the normalization for changes in the dark
matter properties.  The density contrast $(\delta\rho/\rho)_M^2\propto
k^3P(k)^2$ is nearly constant at small mass for CDM-like power
spectra, so that all perturbations collapse at the same time, as
discussed above in \S\ref{sec:hierarchy}.  Changing the mass scale of
the first collapsed objects will therefore not change the initial
epoch of structure formation, and will leave the normalization of $Q$
vs $\sigma$ unchanged, provided that the masses of the new seeds are
still in the regime where $(\delta\rho/\rho)_M^2$ is constant.  If the
core phase space densities of these first small halos are limited by
the primordial phase space density $Q_0$, then we also expect no shift
in the $Q$-$\sigma$ normalization -- at small mass, the filtering mass
scale varies like $M_{filter}\propto Q_0^{-1}$.

On the other hand, if the filtering is on a scale where
$(\delta\rho/\rho)^2$ decreases with $M$, then the first collapsed
objects have both larger masses and smaller primordial phase
densities.  In this case, for a given normalization of the power
spectrum on large scales, the first collapse is sufficiently delayed
such that $Q$ for the first objects drops faster than $M^{-1}$,
reducing the normalization of the $Q$-$\sigma$ relation.  Filtering on
these larger mass scales may also change the expected degree of
scatter in the $Q$-$\sigma$ relation; with additional observational
and theoretical investigation, one could potentially use the
slope and scatter of the $Q$-$\sigma$ relation to place limits on
the mass scale of the first objects.  However, stronger constraints
will probably come from considering the abundance of low mass halos.

\subsection{Links to Elliptical Galaxies}

We note that qualitatively similar behavior to Figure~\ref{fig:Q_vs_V}
is seen within other collisionless systems, namely giant elliptical
galaxies, where the dynamics are dominated by stars within the
half-light radius.  Like dark matter halos, these ellipticals may also
have formed via successive mergers of largely collisionless systems
(particularly in clusters, where progenitors are largely gas-poor).
Paralleling early work by Carlberg (1986) and Lake (1989), Hernquist
et al.\ (1993) used data on elliptical galaxies from Bender et al
(1992) to show a systematic decrease in elliptical galaxy phase
density with increasing luminosity (roughly $Q\propto L_{B}^{-1.5}$).
Assuming a Faber-Jackson (1976) relationship of $L_B \propto
\sigma_0^{4}$, the central phase space densities of ellipticals must
scale with the central velocity dispersion as roughly $Q \propto
\sigma_0^{-\nu}$, with $\nu\sim6$.

The true velocity scaling may be shallower than implied by Hernquist
et al.\ (1993), however.  Recent analyses of elliptical galaxies'
light profiles suggest that ellipticals are non-homologous on large
scales, such that they differ from deVaucouleurs profiles
systematically with increasing mass (Caon et al.\ 1993, Graham \&
Colless 1997), such that lower mass ellipticals have shallower inner
profiles\footnote{Note that we are referring to light profiles and
  phase densities measured on the scale of the half light radii, not
  at the innermost points measured by HST, where the densities may be
  substantially affected by the presence of central black holes.};
Hjorth \& Madsen (1995) argue that taking this non-homology into
account should lead to a shallower relationship between $Q$ and $L$,
and thus the scaling of $Q$ with $\sigma_0$ should be likewise
shallower ($\nu<6$), and possibly compatible with the $\nu\sim3-4$
behavior seen in Figure~\ref{fig:Q_vs_V}.  If a proper reanalysis
shows that the true relationship is steeper than observed in dark
matter halos, then the difference is likely to reflect a more
disruptive, less quiescent merger history offering more opportunities
for breaking homology.

Given the similarities of the global phase space density behavior
shared by both dark matter halos and ellipticals (particularly in the
stellar-dominated inner regions), they may also share the more
detailed internal phase space evolution.  It is plausible that
even with their possibly quieter hierarchy, dark matter halos may
eventually reveal the same homology breaking seen in ellipticals.  The
non-homology in elliptical galaxy light profiles is typically
parameterized in terms of the S\'ersic function, with surface
brightness $\Sigma(r)\propto\exp{[-(r/r_0)^{1/n}]} $ (where $n=4$
corresponds to a deVaucouleurs' profile).  Work by Caon et al.\ (1993)
and Graham \& Colless (1997) shows that the S\'ersic index $n$ tends
to increase systematically with the projected half-light radius $r_e$.
The S\'ersic index $n$ is 1--2 for the smallest ellipticals
($r_e\lesssim 1\kpc$), and increases systematically to 5--10 for the
largest ellipticals ($r_e\sim 10\kpc$). In Figure~\ref{fig:sersic} we
plot the corresponding density profiles for $n=1-7$, using M\'arquez et
al.\ (2000)'s fitting formula for the deprojected density.  The lower
mass ellipticals with $n\sim 1 $ have density profiles with shallower
cores than the higher mass ellipticals (although the overall density
is higher).  This may be analogous to the detection of relatively
shallow inner profiles in the cores of rotating dwarfs and steeper
profiles in massive clusters (although see caveats in
\S\ref{sec:clust}).  We note also that the density profile which
corresponds to a projected deVaucouleur's profile ($n=4$) is quite
similar to the density profile found by Navarro et al.\ (1996) for
simulated dark matter halos, both plotted in
Figure~\ref{fig:sersic_deV}.


\subsection{Constraints on the Mass of Dark Matter Candidates}

In addition to giving us clues about the relaxation processes involved
in galaxy formation, the above observations of the phase space density
can be used to place strong limits on the masses of possible particle
dark matter candidates.  If dark matter has a primordial velocity
dispersion, then its initial phase space density is lowered over CDM,
which presumably leads to lower overall densities in the final
virialized halo.  Generalizing the Tremaine \& Gunn (1979) argument
for massive neutrinos (see also Gerhard \& Spergel  1992),   Hogan \& Dalcanton
(2000) showed that the primordial phase density
$Q_0$ of dark matter particles $X$ can be simply related to the mass
$m_X$ (of particles which decouple when their momentum
distribution is relativistic) through
\begin{equation}      \label{eqn:Q_mx}
Q_0 = q_X g_X m_X^4
\end{equation}
where the coefficient  from the distribution function integral $q_X$ is 0.00196
for thermal particles and 0.0363 for degenerate fermions, and $g_X$ is the
number of effective photon degrees of freedom of the particle $X$.  

In the absence of dissipation, the coarse-grained phase space
density can only decrease from its primordial value.  The maximum
observed phase space density therefore places a lower limit on the
mass of the $X$ particle.  Figure~\ref{fig:Q_vs_V} shows that the highest
observed phase space densities are found for dwarf spheroidals, with
$Q_{obs}\!\sim\!10^{-4}\,\msun\pc^3(\kms)^{-3}$.  This lower limit on
$Q_0$ implies that
\begin{equation}
m_X > 669\eV \, 
        \left(\frac{Q_{obs}}{10^{-4}\,\msun\pc^{-3}(\kms)^{-3}}\right)^{1/4}
        \left(\frac{0.00196}{q_X}\right)^{1/4}
        \left(\frac{2}{g_X}\right)^{1/4}
\end{equation}
For thermal particles with 2 degrees of freedom, the data 
suggest that to
$m_X > 669eV$.  For degenerate fermions, with 2 degrees of freedom,
$m_X > 322\eV$. 

These lower limits on particle mass correspond to the largest $Q$
actually observed. There may well be dark matter halos with smaller
velocity dispersions and larger $Q$ halos; indeed, these are expected
in CDM.  However, such systems would never form stars, and thus would
remain undetected.  At velocity dispersions below 7 km/sec, the
collisional cooling of zero-metal atomic gas becomes very inefficient.
The specific binding energy of the shallowest observed halo potentials
is close to to the minimum temperature expected for the protogalactic
medium during early galaxy formation.

For standard collisionless warm matter, if the dark matter saturates
the $Q_0$ limit from dwarf spheroidals (that is, $m_X\approx 700$ eV),
there is a corresponding filtering scale at the masses of dwarf
galaxies (see Sommer-Larsen and Dolgov 1999, Hogan and Dalcanton
2000).  This effect may already be indicated by the paucity of dwarf
galaxies relative to standard CDM predictions.  We do not yet know
enough about the predictions of WDM to compare with halo mass
functions in detail, however.

\section{Conclusion}                    \label{sec:conclusion}

The above examinations of the existing data on the structure and phase
space density of dark matter halos yields a number of conclusions
which may be relevant to constraining the nature of the dark matter.

First, the behavior of halo core size with increasing mass suggests
that it is unlikely that either phase-space packing or highly
collisional dark matter is sufficient for simultaneously explaining
the dark matter cores of dwarf spheroidals, rotating dwarf galaxies,
and clusters of galaxies.  The generic
$r_{core}\propto1/\sqrt{\sigma}$ behavior of these scenarios for core
formation would predict far larger cores for the dwarf spheroidals
than for larger systems, in contrast to observational evidence.

Second, if there were still any doubt, the dramatic decrease in the
characteristic phase space density with increasing mass is extremely
strong evidence for a ``bottom-up'' hierarchical buildup of bound
structures.  Given that the phase space density can
never increase with successive mergers (in the absence of
dissipation), smaller structures cannot generically have fragmented
from larger ones. The trend points to dwarf spheroidals as the lowest
entropy, and therefore dynamically most primitive observable systems.

Third, the observed dependence of $Q\propto \sigma^{-3}$ approximately
agrees (over eight orders of magnitude in $Q$!) with a simple scaling
relation that assumes the minimal decrease compatible with a gentle
merging hierarchy with virial equilibrium and homologous tidal stripping
at each stage.  The same scaling is also compatible with a simple
synchronous collapse of different mass systems to constant virial
density. The two descriptions are both appropriate, at different
stages, for the collisionless hierarchy predicted in CDM models.
We also discuss the physics which sets the normalization of
the $Q$-$\sigma$ scaling.

Fourth, examination of the phase space density for dark matter halos
suggests some parallels to elliptical galaxies.  In most
scenarios, both dark matter halos and cluster elliptical galaxies are
thought to be formed through collisionless merging and accretion.  We
show that observationally, both systems show similar decreases in the
coarse-grained phase space density with increasing velocity
dispersion.  They also show surprisingly similar density profiles.  If
elliptical galaxies can be used as a rough analog to dark matter
profiles, then they suggest that the structure of dark matter halos
may undergo subtle, systematic deviations from homology, leading to
somewhat flatter inner cores (and steeper fall-off at large radii) for
low mass halos.  The amplitude of these deviations are sufficiently
small that they are unlikely to be well resolved in current numerical
simulations, though they may have already been detected
observationally. This non-homology could help to alleviate some of the
discrepancies between observations of rotating dwarfs and predictions
of dark matter simulations on the smallest, most poorly resolved
scales.  This parallel also provides suggestive, although not
conclusive, evidence that dark matter halos are indeed collisionless
on the scale of halo cores.

Fifth, the very high phase space densities of dwarf spheroidals can be
used to place constraints on the masses of potential dark matter
candidates.  For dark matter particles with 2 degrees of freedom,
decoupled while still relativistic, masses of $m_X > 700\eV$ (thermal
fermions) or $m_X > 300\eV$ (degenerate fermions) are preferred.
Because systems with smaller masses and higher phase space densities
than dwarf spheroidals may exist, the actual particle masses may be
substantially higher than these limits. This seems likely since
smaller $\sigma$ halos would be invisible even if they exist; in other
words we see halos populated with stars right up to the cooling limit,
which would be a coincidence if they also correspond to the phase
density limit. This interpretation is again consistent with the view
that dwarf spheroidals are the most primitive bound systems so far
observed.

Finally, we conjecture that the addition of primordial velocity
dispersion can help to reconcile the discrepancies between numerical
predictions of dense central cores in hierarchical clustering (e.g.\ 
Navarro et al.\ 1996,1997, Fukushige \& Makino 1997, Moore et al.\ 
1998\&1999, Ghinga et al.\ 1999, Jing \& Suto 2000) and observations of much
lower central densities (e.g.\ Flores \& Primack 1994, Moore 1994,
Burkert 1995, Navarro et al.\ 1996, Stil 1999).  Regardless of the
origin of the ``universal'' density profile, simulations routinely
predict a denser, more concentrated halo than is actually observed; in
other words, the predicted central phase space density for CDM is too
high, which in turn suggests that the CDM initial conditions
themselves have too high a phase density.  Instead, if the primordial
phase space density is lower than the CDM case, then even if the final
density structure is set by merging and relaxation, the final phase
space density should be lowered as well, provided the primordial $Q_0$
is not much higher than that which occurs as a result of virialization
at the earliest nonlinear collapse in CDM.  Numerical simulations by
Huss et al.\ (1999) approximately explore this conjecture through
studying monolithic collapse of spherical overdensities with varying
velocity dispersion.  However, numerical relaxation is clearly a
problem for these simulations, and we draw no conclusion from them at
this time.  Future numerical work will certainly shed light on this
hypothesis.

\acknowledgements

We are happy to acknowledge discussions with Tom Quinn, David Spergel,
Frank van den Bosch, Arif Babul, Simon White, Julio Navarro, George
Lake, and Ben Moore.  JJD gratefully acknowledges the hospitality of
the Institute of Theoretical Physics in Santa Barbara, where some of
this work was done.  The research was supported in part by the
National Science Foundation under Grant No.\ PHY94-07194, and at the
University of Washington by NASA and NSF.

\section{References}

\hi{Allen, S. W. 1998, \mnras, 296, 392}

\hi{Allen, S. W., Fabian, A. C., \& Kneib, J.-P. 1996, \mnras, 279, 615}

\hi{Bender, R., Burstein, D., \& Faber, S. M. 1992, \apj, 399, 462}

\hi{Bento, M. C., Bertolami, O, Rosenfeld, R., \& Teodoro, L. 2000, preprint,
(astro-ph/0003350)}

\hi{Blumenthal, G. R., Faber, S. M., Flores, R., \& Primack, J. R. 1986,
\apj, 301, 27}

\hi{Borriello, A., \& Salucci, P. 2000, \mnras, submitted (astro-ph/0001082)}

\hi{Burkert, A. 1995, \apjl, 447, L25}

\hi{Burkert, A. 1997, \apjl, 474, L99}

\hi{Burkert, A. 2000, \apjl, submitted (astro-ph/0002409)}

\hi{Caon, N, Capaccioli, M, \& D'Onofrio, M. 1993, \mnras, 265, 1013}

\hi{Carignan, C., \& Beaulieu, S. 1989, \apj, 347, 760}

\hi{Carignan, C., \& Puche, D. 1990, \aj, 100, 641}

\hi{Carlberg, R. G. 1986, \apj, 310, 593}

\hi{Carlberg, R. G., Yee, H. K. C., Ellingson, E., Morris, S. L., Abraham,
R. , Gravel, P., Pritchet, C. J., Smecker-Hane, T., Hartwick, F. D. A.,
Hesser, J. E., Hutchings, J. B., \& Oke, J. B. 1997, \apjl, 485, L13}

\hi{Colin, P., Avila-Reese, V., \& Valenzuela, O. 2000, 
\apj, submitted (astro-ph/0004115)}

\hi{Dalcanton, J. J., \& Bernstein, R. A. 2000, in {\it XVth IAP
    Meeting Dynamics of Galaxies: from the Early Universe to the
    Present}, ed. F. Combes, G. A. Mamon, \& V. Charmandaris}

\hi{Ettori, S., \& Fabian, A. C. 1999, \mnras, 305, 834}

\hi{Firmani, C., D'Onghia, E., Aliva-Reese, V., Chincarini, G., \& Herna\'ndez,
X. 2000, \mnras,, in press}

\hi{Flores, R. A., \& Primack, J. R. 1994, \apjl, 427, L1}

\hi{Flores, R. A., \& Primack, J. R. 1996, \apjl, 457, L5}

\hi{Fukushige, T., \& Makino, J. 1997 \apj, 477, L9}

\hi{Gerhard, O. E. \& Spergel, D. N. 1992, \aj, 389, L9}

\hi{Goodman, J. 2000, preprint, (astro-ph/0003018)}

\hi{Graham, A., \& Colless, M. 1997, \mnras, 287, 221}

\hi{Hannestad, S., \& Scherrer, R. J. 2000, preprint, (astro-ph/0003046)}

\hi{Hannestad, S., 1999, astro-ph/9912558}

\hi{Hjorth, J., \& Madsen, J. 1995, \apj, 445, 55}

\hi{Hernquist, L., Spergel, D. N., \& Heyl, J. S. 1993, \apj, 416, 415}

\hi{Hogan, C. J. 1999, \apj, 527, 42}

\hi{Hogan, C. J., \& Dalcanton, J. J. 2000, Phys. Rev. D, in press}

\hi{Hu, W., Barkana, R., \& Gruzinov, A. 2000, \prl, submitted, (astro-ph/0003365)}

\hi{Irwin, M., \& Hatzidimitriou, D. 1995, \mnras, 277, 1354}

\hi{Jing, Y. P. \& Suto Y. 2000, \apj, 529, 69}

\hi{Jobin, M., \& Carignan, C. 1990, \aj, 100, 648}

\hi{Kamionkowski, M., \& Liddle, A. R. 1999, preprint, (astro-ph/9911103)}

\hi{Klenya, J., Geller, M., Kenyon, S., \& Kurtz, M. 1999, \aj, 117, 1275}

\hi{Kravtsov, A. V., Klypin, A. A., Bullock, J. S., \& Primack,
  J. R. 1998, \apj, 502, 48}

\hi{Kochanek, C. S., \& White, M. 2000, \apj, submitted, (astro-ph/0003483)}

\hi{Kuhn, J. R., \& Miller, R. H. 1989, \apj, 341, L41}

\hi{Kuhn, J. R., Smith, H. A. \& Hawley, S. L. 1996, \apj, 469. L93}

\hi{Lake, G. 1989, \aj, 97, 1312}

\hi{Lake, G. 1989, \aj, 98, 1253}

\hi{Majewski, S. R., Ostheimer, J. C., Patterson, R. J., Kunkel, W. E.,
Johnston, K. V., \& Geisler, D. 2000, \apj, 119, 760}

\hi{Mateo, M. 1998, \araa, 36, 435}

\hi{Mateo, M., Olszewski, E. W., Vogt, S. S., \& Keane, M. J. 1998, 116, 2315}

\hi{Mellier, Y., Fort, B., \& Kneib, J.-P. 1993, \apj, 407, 33}

\hi{Mohapatra, R. N., \& Teplitz, V. L. 2000, preprint, (astro-ph/0001362)}

\hi{Moore, B., 1994, \nat, 370, 629}

\hi{Moore, B., Quinn, T., Governato, F., Stadel, J., \& Lake, G. 1999,
\mnras, 310, 1147}

\hi{Moore, B., Governato, F., Quinn, T., Stadel, J., \& Lake, G. 1998,
\apj, 499, L5}

\hi{Moore, B., Gelato, S., Jenkins, A., Pearce, F. R., \& Quilis, V. 2000, 
preprint, (astro-ph/002308)}

\hi{Navarro, J. F. 1998, preprint, (astro-ph/9807084)}

\hi{Navarro, J. F., Frenk, C. S., \& White, S. D. M. 1996, \apj, 462, 563}


\hi{Navarro, J. F., Frenk, C. S., \& White, S. D. M. 1997, \apj, 490, 493}

\hi{Oh, K. S., Lin., D. N. C., \& Aarseth, S. J. 1992, \apj, 442, 142}

\hi{Olling, R. P. 1995, \aj, 110, 591}

\hi{Olling, R. P., \& Merrifield, M. R. 1999, \mnras, in press}

\hi{Peebles, P. J. E. 2000, preprint, (astro-ph/0002495)}

\hi{Pierre, M., Le Borgne, J. F., Soucaile, G., \& Kneib, J.-P. 1996, \aa,
        311, 413}

\hi{Pryor, C., \& Kormendy, J. 1990, \aj, 100, 127}

\hi{Riotto, A. \& Tkachev, I. 2000, preprint, (astro-ph/0003388)}

\hi{Sackett, P. D. 1999, in {\it Galaxy Dynamics}, eds.\ Merritt, D.,
        Sellwood, J. A., \& Valluri, M., ASP}

\hi{Sellwood, J. A. 2000, \apjl, submitted, (astro-ph/0004352)}

\hi{Sommer-Larsen, J., and Dolgov, A. 1999, preprint, (astro-ph/9912166)}

\hi{Spergel, D. N., \& Steinhardt, P. J. 1999, \prl, submitted
(astro-ph/9909386)}

\hi{Schwarz, D. J., \& Hofmann, S. 1999, preprint, (astro-ph/9912343)}

\hi{Shi, X., and Fuller, G. M. 1999, \prl, 82, 2832}

\hi{Swaters, R. 2000, Ph.D.\ thesis, University of Groningen}

\hi{Syer, D., \& White, S. D. M. 1998, \mnras, 293, 337}

\hi{Tremaine, S. and Gunn, J. E. 1979, Phys.\ Rev.\ Lett., 42, 407}

\hi{Tyson, J. A., Kochanski, G. P., \& Dell'Antonio, I. P. 1998, \apj,
        498, L107}

\hi{van den Bosch, F. C., Robertson, B. E., Dalcanton, J. J., \& de Blok, W. J. G. 2000, \aj, 119, 1579}

\hi{van den Bosch, F. C., \& Swaters, R. 2000, \aj, submitted}

\hi{van der Marel, R. P. et al. 2000, \aj, in press (astro-ph/9910494)}

\hi{Williams, L. L. R., Navarro, J. F., \& Bartelmann, M. 1999, \apj, 527, 535}

\hi{Yoshida, N., Springel, V., White, S. D. M., \& Tormen, preprint,
  (astro-ph/0002362)}

\vfill
\break
\section{Figure Captions}

\figcaption[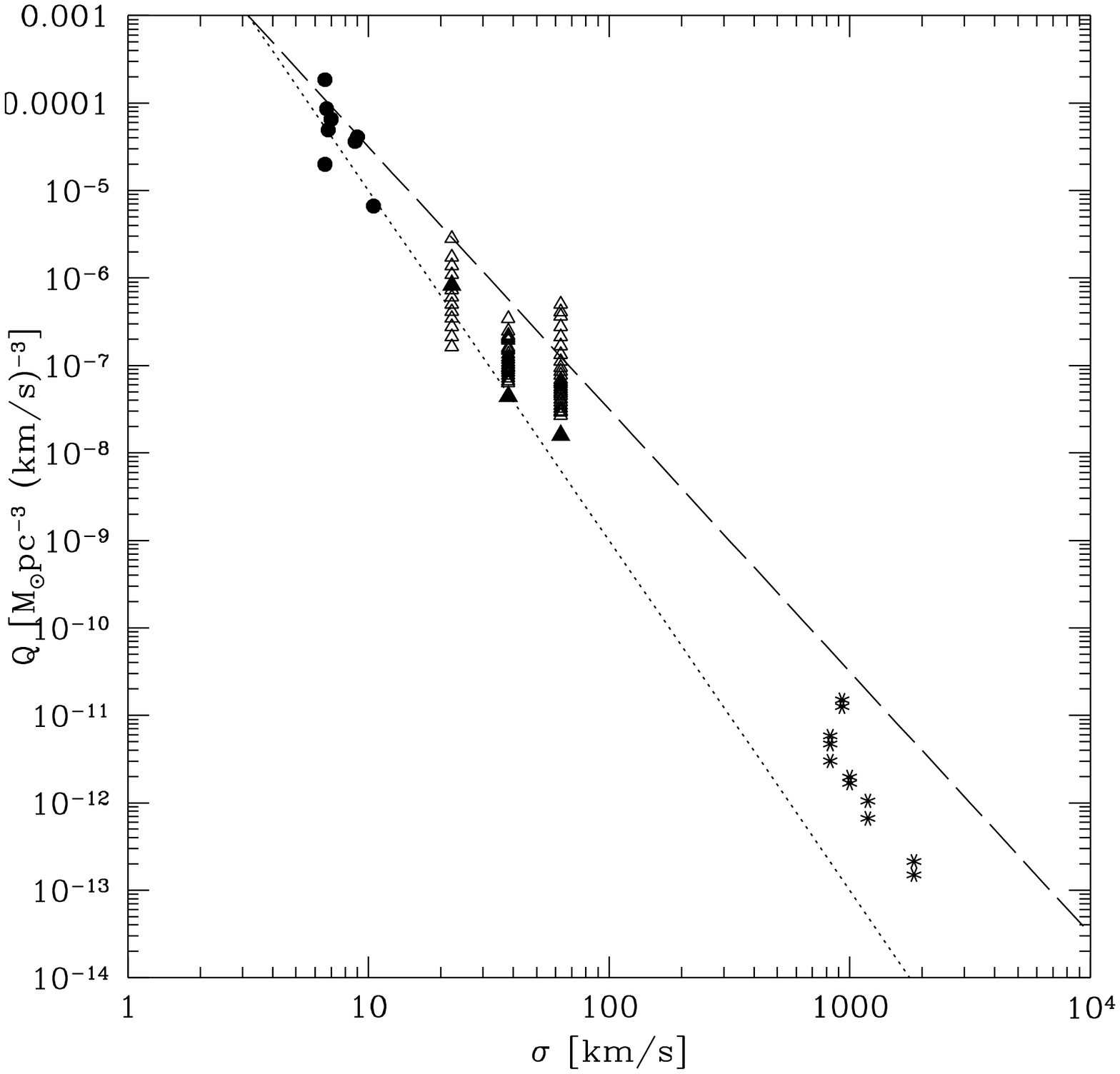]{ Mean interior phase space density $Q$ within the approximate
core radius as a function of the velocity dispersion of the system.
Circles = dwarf spheroidals, Triangles = rotating dwarfs (DDO 154, NGC
247, and NCG 3109 in order of decreasing maximum $Q$; innermost points
have highest $Q$, and solid triangle marks the phase density at the
core radius), Asterix = clusters (multiple points at the same velocity
dispersion represent different mass determinations for the same
cluster, given within the radius of a strongly lensed arc).  Dashed
line shows $Q\propto\sigma^{-3}$ scaling, the minimal predicted
decrease in $Q$.  Dotted line shows $Q\propto\sigma^{-4}$, for reference.
\label{fig:Q_vs_V}}

\figcaption[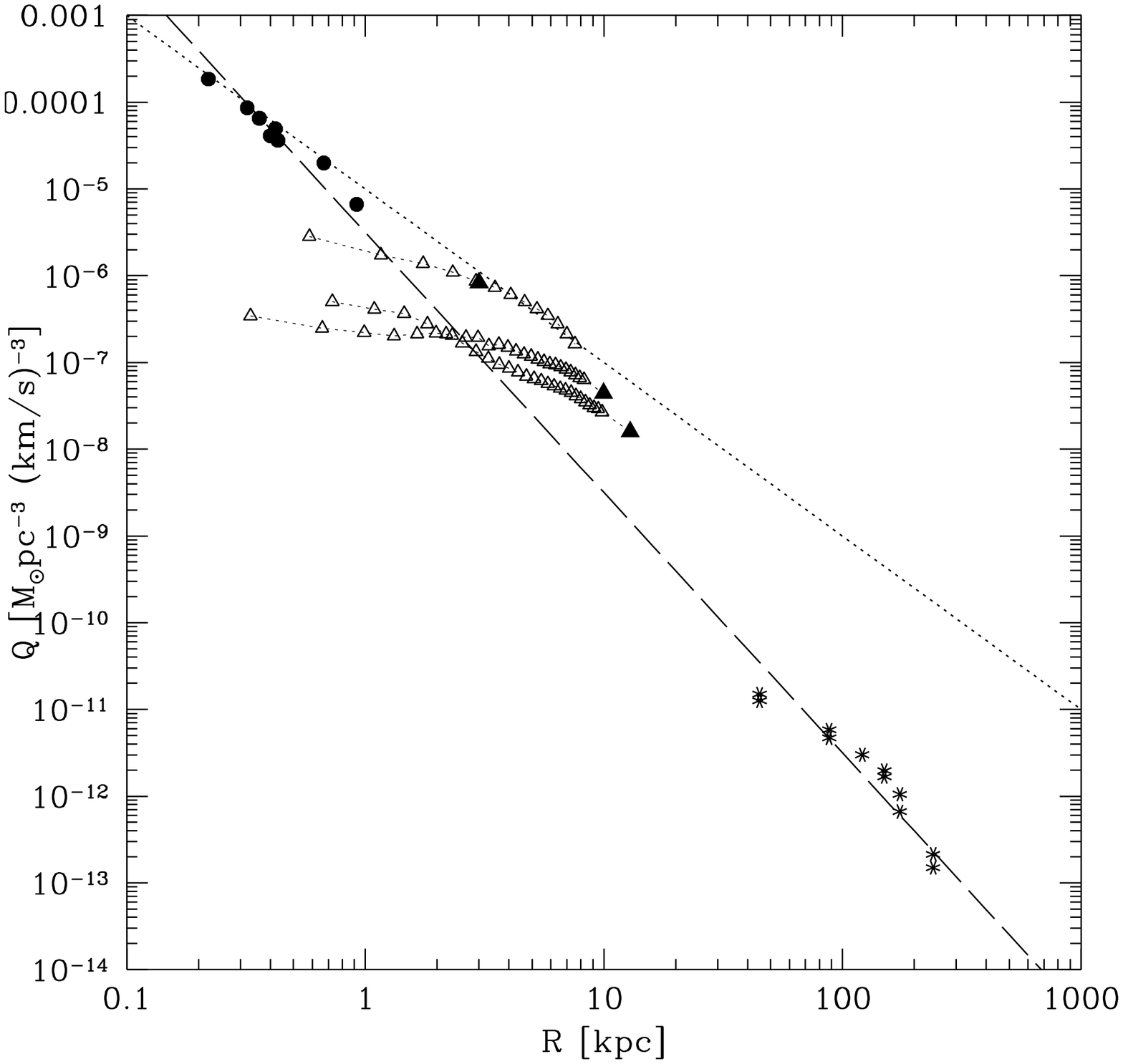]{ Mean interior phase space density $Q$ as
  a function of the radius within which Q was measured.  Circles =
  dwarf spheroidals, Triangles = rotating dwarfs (DDO 154, NGC 247,
  and NCG 3109 in order of decreasing maximum $Q$; solid triangle
  marks the phase density at the core radius), Asterix = clusters
  (multiple points at the same radius represent different mass
  determinations for the same cluster, given within the radius of a
  strongly lensed arc).  The apparent correlation between $Q$ and
  $r_{core}$ for dwarf spheroidals results from the lack of
  significant variation in the velocity dispersion of the dwarf
  spheroidals, such that the variation in $Q$ is driven entirely by
  the variation in $r_{core}$; we consider the uncertainties in
  $r_{core}$ to be sufficiently large that this apparent correlation
  is not necessarily physically meaningful.  Dashed line shows
  $Q\propto R^{-3}$ scaling, the minimal predicted decrease in $Q$.
  Dotted line shows $Q\propto R^{-2}$, for reference.
\label{fig:Q_vs_R}}

\figcaption[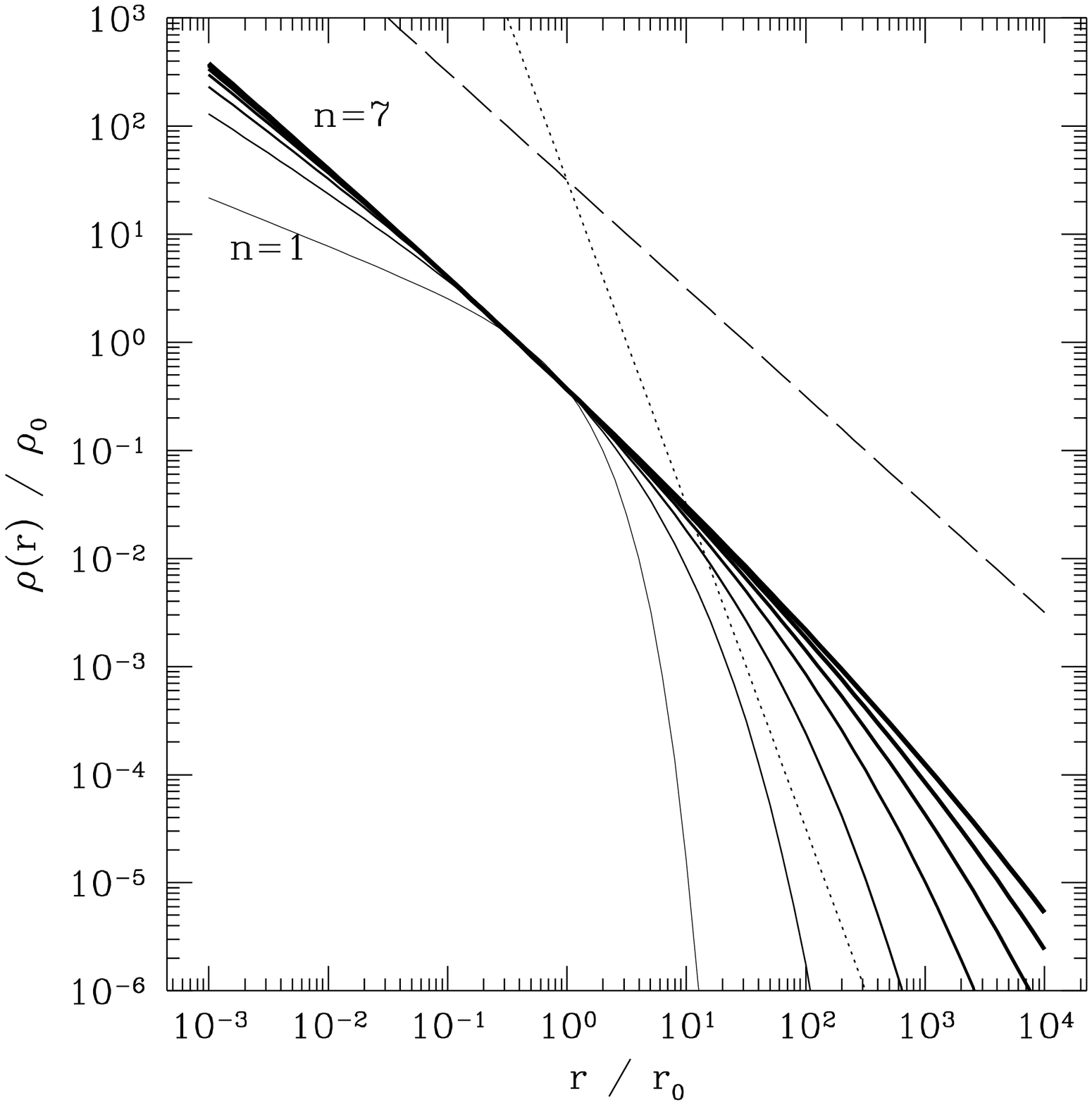]{ Density profiles corresponding to
  projected S\'ersic surface density profiles
  ($\Sigma(r)\propto\exp{[-(r/r_0)^{1/n}]}$) with $n=1-7$ (larger $n$
  = heavier line weight).  The dashed and dotted lines represent
  $\rho\propto r^{-1}$ and $\rho\propto r^{-3}$, respectively.
\label{fig:sersic}}

\figcaption[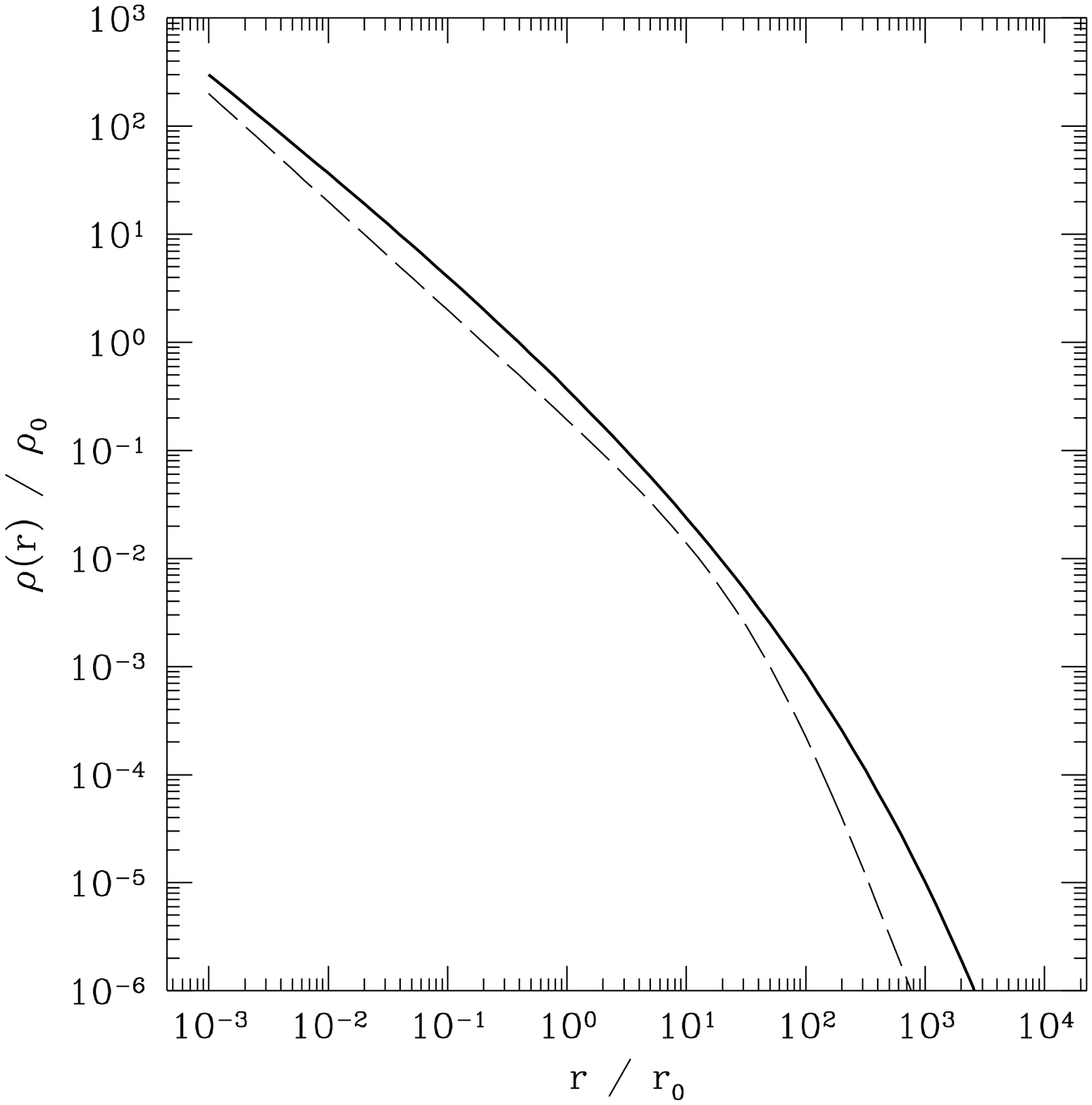]{ Density profile of an $n=4$ S\'ersic
  profile (solid line), compared to an NFW density profile with
  $\rho\propto (r/a)^{-1}(1+r/a)^{-2}$, with $a=50r_0$ (dashed line).
\label{fig:sersic_deV}}

\vfill
\clearpage


\ifprintfig

\begin{figure}[p]
\centerline{ \psfig{figure=Dalcanton.f1.ps,height=7.5in} }
\begin{flushright}{\bigskip\cap Figure \ref{fig:Q_vs_V}}\end{flushright}
\end{figure}
\vfill
\clearpage
\begin{figure}[p]
\centerline{ \psfig{figure=Dalcanton.f2.ps,height=7.5in} }
\begin{flushright}{\bigskip\cap Figure \ref{fig:Q_vs_R}}\end{flushright}
\end{figure}
\vfill
\clearpage
\begin{figure}[p]
\centerline{ \psfig{figure=Dalcanton.f3.ps,height=7.5in} }
\begin{flushright}{\bigskip\cap Figure \ref{fig:sersic}}\end{flushright}
\end{figure}
\vfill
\clearpage
\begin{figure}[p]
\centerline{ \psfig{figure=Dalcanton.f4.ps,height=7.5in} }
\begin{flushright}{\bigskip\cap Figure \ref{fig:sersic_deV}}\end{flushright}
\end{figure}
\vfill
\clearpage
\fi

\end{document}